\documentclass{optica-article}

\journal{opticajournal} 

\articletype{Research Article}
\usepackage{xcolor}
\usepackage{lineno}
\usepackage{float}

\usepackage{ulem}  
\usepackage{comment} 

\usepackage{diagbox} 
\definecolor{green1}{RGB}{0, 125, 0} 
\definecolor{green2}{RGB}{144, 200, 144} 
\nolinenumbers 
\usepackage{xr}

\begin{document}

\title{Modified Transfer Matrix Method for the Extraction of Material Properties via Terahertz Time-Domain Spectroscopy}

\author{Kamyar Rashidi\authormark{1,2} and Matthew Y. Sfeir\authormark{1,2,*}}

\address{
\authormark{1}Photonics Initiative, Advanced Science Research Center, City University of New York, New York, NY 10031, USA\\
\authormark{2}Department of Physics, Graduate Center, City University of New York, New York, NY 10016, USA
}

\email{\authormark{*}msfeir@gc.cuny.edu}

\begin{abstract*} 
Terahertz Time-Domain Spectroscopy is a powerful technique for extracting the low-frequency optical properties of materials. However, the optical constants are difficult to determine directly from the experimental transfer function, such that various numerical approximations must be implemented to describe specific conditions. Here, we introduce a modified Transfer Matrix Method that uses a genetic algorithm for optimization to determine the refractive index of materials in the THz regime. We show that this approach is generally applicable across a wide range of refractive indices, structures, and frequency ranges. Our method is intuitive and yields accurate results compared to leading methods across a wide spectral range (0.1 -- 15 THz).

\end{abstract*}

\section{Introduction}

Terahertz time-domain spectroscopy (THz-TDS), prized for its high signal-to-noise ratio and the comprehensive data captured from time-domain sampling of THz waveforms, is widely used across multiple disciplines, including medical imaging \cite{sun2017recent, stantchev2020real}, chemical detection \cite{fischer2005chemical}, environmental monitoring \cite{prokscha2025perspectives}, terahertz communication systems \cite{kleine2011review},  and investigations of phonon modes and material conductivity \cite{thorsmolle2004ultrafast, yang2016biomedical}.
Recent advancements in emitter and detector technologies has extended the frequency range of THz-TDS to $\sim$ 40 THz  \cite{seifert2016efficient,müller2020phase}, enhancing its material characterization capabilities beyond the range of conventional THz systems
 \cite{zhang2016broadband}.
This technique has long been an established method for determining the complex refractive index, conductivity, and thickness of materials from time-domain traces\cite{peretti2018thz, ulatowski2020terahertz}. Its key advantage lies in its ability to directly measure the THz electric field with \textcolor{black}{femtosecond} time resolution, rather than focusing solely on the averaged energy. This enables access to both the real and imaginary components of the refractive index, without relying on the Kramers-Kronig relations\cite{Nuss1998}. To determine  the refractive index of the sample in THz-TDS, the parameters in a \textcolor{black}{transfer function model} are optimized to fit the experimental transfer function. However, selecting an appropriate transfer function model and optimization algorithm can be challenging, particularly when aiming for convergence to a physically meaningful and accurate refractive index \cite{tayvah2021nelly,jepsen2019phase}.

\textcolor{black}{Several transfer function models have been proposed to describe light propagation though a sample that are tailored for specific experimental conditions, including the number, thicknesses, and refractive index of the layers, as well as the key details such as the time window over which the waveform is measured\cite{withayachumnankul2008uncertainty, greenall2016multilayer,cassar2019iterative, whelan2020reference, neu2018applicability}. In principle, all one-dimensional propagation could be described with Transfer Matrix Method (TMM) approaches, which accounts for all reflections and transmissions inside a multilayer structure \cite{troparevsky2010transfer,waddie2020terahertz,rashidi2024efficient,michail2024addressing,katsidis2002general}. However, its applicability in THz-TDS is limited to very thin layers where all reflections fall within the measured time window. For thicker and more complex structures, TMM accounts for propagation paths that are beyond the measured time window, resulting in calculated refractive indices that diverge from the actual value. To address this issue, several approximations are implemented to modify the TMM approach. The simplest modification ignores all reflections, considering only the propagation across an interface between two adjacent layers and the propagation within each layer\cite{withayachumnankul2014fundamentals}. We refer to this approach here as the single-pass method (SPM). While this is a reasonable approximation for thick layers, whose reflections occur outside the measured time window, low index layers with thicknesses up to several hundred nanometers will still have reflections that occur within the relevant time window and cannot be ignored \cite{withayachumnankul2008uncertainty}. 
For a combination of thin and thick materials, a hybrid approach has been proposed that integrates Fabry-Perot terms for thin layers and a single transmission term for thick layers. However, this approach remains an approximation and presents significant challenges when addressing multiple closely spaced thin films \cite{neu2018applicability,peretti2018thz}.
Another useful approach uses a tree algorithm \cite{cassar2019iterative, greenall2016multilayer} to identify components resulting from multiple reflection  with net amplitudes that exceed the instrument's detection limit \cite{tayvah2021nelly}. This method is the basis for Nelly, a freely-available software package for the Matlab numerical computing environment\cite{tayvah2021nelly}. However, this approach becomes difficult to implement in complex multilayer structures since the number of potential reflections that must be considered increases exponentially with number of interfaces. \cite{cassar2019iterative,tayvah2021nelly,wilk2008highly,greenall2016multilayer}. Furthermore, this approach ignores paths with smaller amplitudes that accumulate and surpass, in total, the instrument's threshold.} As a result, a general, broadly applicable method is still lacking.

\textcolor{black}{In addition to the transfer function model, a suitable analytical method 
 or numerical algorithm is necessary to obtain good convergence to the experimental data while avoiding nonphysical solutions. 
Analytical methods are intuitive and faster than numerical methods when a unique solution exists, and can be used to validate numerical approaches. However, in practice, they can only be used for THz-TDS in cases where overly simplistic assumptions are used, such as ignoring multiple reflections (e.g. in the SPM model) and considering only materials with very low absorption. These analytical methods fail if the underlying assumptions do not hold \cite{duvillaret1999highly}. The advent of high-performance computing and advanced optimization methods enables us to accurately finds the complex refractive index that minimizes the deviation between the modeled transfer function and the experimental transfer function at each frequency point. Common optimization algorithms are the gradient-based method \cite{ma2020application} or Nelder--Mead simplex method \cite{lagarias1998convergence, tayvah2021nelly, pupeza2007highly}, where, given the thickness of the sample, the refractive index at each wavelength is optimized separately to match the experimental transfer function. While these methods offer speed, they are limited by their tendency to converge to local minima specific to a given set of guess solutions\cite{peretti2018thz,chehouri2015review}. 
An alternative is the use of artificial neural networks, which can model the relationship between the input training data and the corresponding output by adjusting the weights and biases of the connections between neurons. However, neural networks require prior knowledge of the sample's refractive index through multiple datasets for effective training \cite{klokkou2022artificial,mirjalili2014grey,Beddoes25}, which are currently difficult to obtain. To address the challenges of complex problems with multiple local minima,  population-based metaheuristic optimizations have been developed. Unlike machine learning approaches, no training set or prior knowledge of the sample is required. The advantage of this approach is that they explore a larger portion of the search space and provide a range of possible refractive indices that are independent of initial conditions\cite{kennedy1995particle,mirjalili2014grey,rashidi2023design, rashidi2018optimal}. This is especially important in THz-TDS, in which solutions to the refractive index may include multiple integers of the phase of the transfer function\cite{jepsen2019phase}. }
 
Here, we present an approach based on a newly formulated transfer function model, combined with an advanced optimization strategy that allows us to determine the complex refractive indices of materials at THz frequencies with reduced complexity and increased accuracy. Our implementation is based on modifications to the Transfer Matrix Method (which we refer to as the modified matrix matrix method or MTMM) that accounts only for multiple reflections that contribute to the measured  signal within the relevant time window, i.e., ignoring reflections in thick substrates. \textcolor{black}{Importantly, we implement metaheuristic optimization techniques to explore a wide search space without falling into local minima, avoiding spurious solutions. To eliminate physically meaningless dispersion curves, we initialize the optimization routine with an analytical approximation and constrain the search space to physically meaningful solutions. Our findings demonstrate that the combination of the MTMM and the use of metaheuristic optimization techniques can address the challenges associated with uncertainty in extracting material properties from THz-TDS.} Source code for the MTMM method is freely available at \url{https://github.com/SfeirLab/MTMM-THz-TDS}.

\section{Methods}

\subsection{Transfer Function Model: Modified Transfer Matrix Method \label{sec: MTMM}}

\begin{figure}
    \centering
\includegraphics[width=1\linewidth]{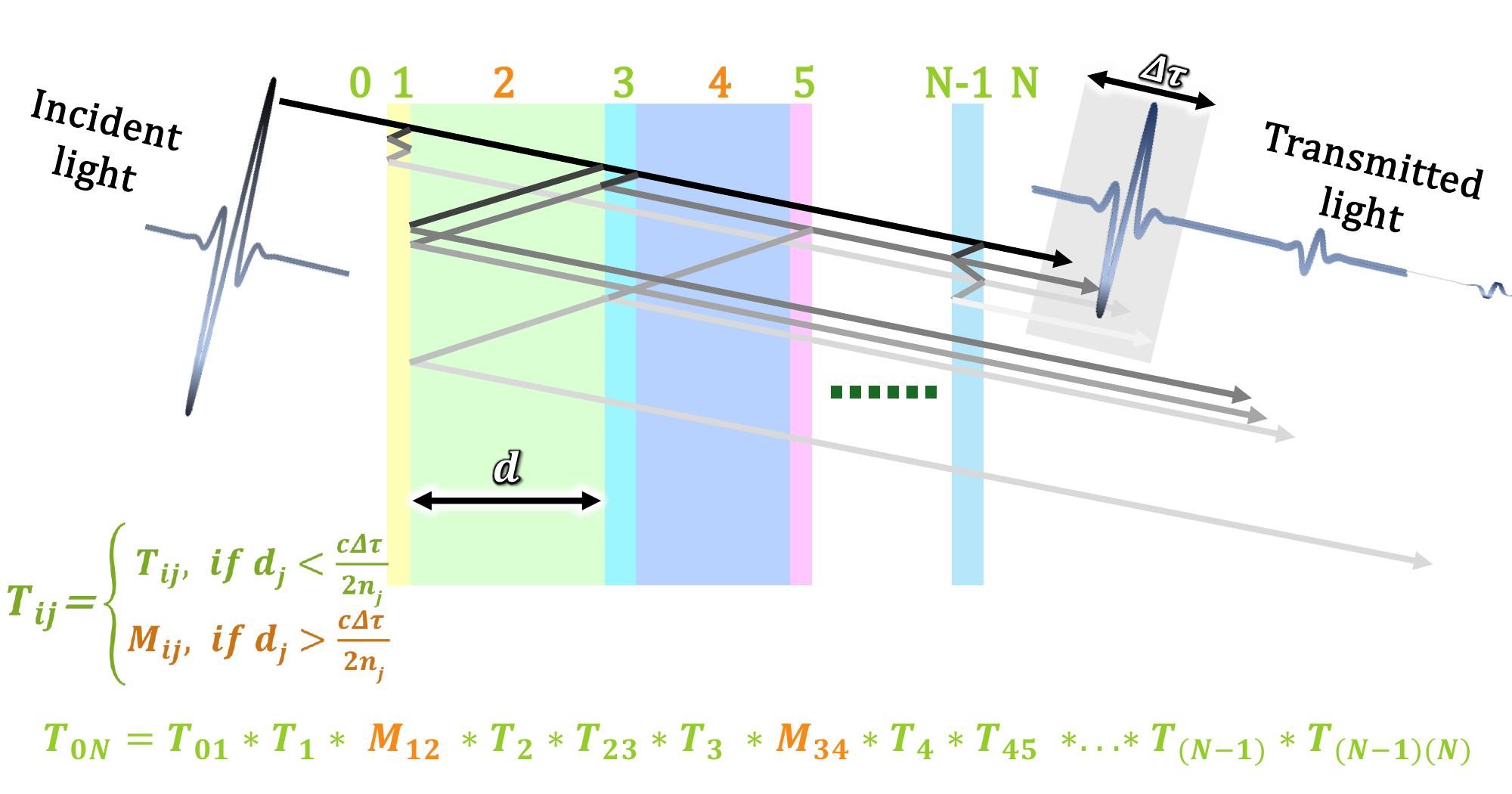} 
\caption{Schematic of electromagnetic wave propagation through an \(N\)-layer system and its transfer function formulation using the modified transfer matrix method (MTMM). The wave is shown at an angle for clarity, although it is normally incident at 0°. The beam splits at each layer interface, and each successive path exhibits decreasing amplitude, with some remaining within the time window ($\Delta \tau$). Paths involving at least one round trip through the thicker layers extend beyond the first window. For these thicker layers, the transfer function is modified to disregard reflected light falling outside the window ($M_{ij}$). In contrast, the unmodified transfer matrices for thinner layers (within the time window) remain the same ($T_{ij}$).
}
  \label{fgr:fig1}
\end{figure}

We have designed MTMM to be a robust transfer function model that uses the full TMM method for thin layers, where multiple reflections within the temporal observation window are included without approximation, while substituting a modified term to exclude multiple reflections in thick components with long temporal delays. Figure~\ref{fgr:fig1} shows a schematic of a typical multilayer structure, highlighting the layers for which the transfer function is refined according to our scheme. Backward-propagating waves from layers exceeding a threshold thickness are omitted from the transfer function, as they arrive outside the defined time window constrained by the delay line length.
The threshold thickness is defined from the measurement window \(\Delta \tau\) and the refractive index \(n_{\text{j}}\) of the \(j\)th layer as:
\begin{equation}
d_{\text{th,j}} = \frac{c \cdot \Delta \tau}{2n_{\text{j}}},
\label{eq:dT}
\end{equation}
where \(c\) is the speed of light in vacuum. In typical THz-TDS measurements, only a single layer has an unknown refractive index. As an estimate, we use the analytical method given by:
\begin{equation}
n_{\text{eff}} = 1 + \frac{c \Delta t}{d},
\label{eq:neff}
\end{equation}
where \(d\) is the thickness of the unknown layer, and \(\Delta t\) is the peak amplitude time delay between a pulse traveling through the reference and one traveling through the sample~\cite{klokkou2022artificial}. Each layer is then evaluated against our threshold criteria.

In the TMM, there is a transfer function associated with each layer and each interface. For normal incidence of a wave passing through the sample, the transfer matrix \( T_{ij} \) for wave propagation across the interface located at \( x_{i} \) between layers \( i \) and \( j \) is given by \cite{yeh1990optical}:

\begin{equation}\label{eq:Tij}
\begin{pmatrix}
E_F(x_{i}^{-}) \\
E_B(x_{i}^{-})
\end{pmatrix}
= T_{ij} \begin{pmatrix}
E_F(x_{i}^{+}) \\
E_B(x_{i}^{+})
\end{pmatrix} =
\begin{pmatrix}
\frac{n_i + n_j}{2 n_i} & \frac{n_i - n_j}{2 n_i} \\
\frac{n_i - n_j}{2 n_i} & \frac{n_i + n_j}{2 n_i}
\end{pmatrix} 
\begin{pmatrix}
E_F(x_{i}^{+}) \\
E_B(x_{i}^{+})
\end{pmatrix}.
\end{equation}
where the subscripts \(F\) and \(B\) denote forward and backward waves, and the superscripts \(-\) and \(+\) indicate just before and just after the interface between layers \( i \) and \( j \) within the multilayer stack. Similarly, the transfer matrix \( T_i \) for wave propagation through layer \( i \) is:
\begin{equation}
T_i =
\begin{pmatrix}
e^{i\Phi_i} & 0 \\
0 & e^{-i\Phi_i}
\end{pmatrix},
\end{equation}
where we have defined \( \Phi_i = k_i d_i \). The complete transfer matrix for a multilayer system is the product of all individual transfer matrices:
\begin{equation}
T_{0N} = T_{01} T_{1} T_{12} T_{2} \ldots T_{(N-1)} T_{(N-1)N}.
\end{equation} and can be used to describe wave propagation from layer 0 to layer \( N \):

\begin{equation}
\begin{pmatrix}
E_F(x_{0}^{-}) \\
E_B(x_{0}^{-})
\end{pmatrix}
=
\begin{pmatrix}
T_{0N}^{11} & T_{0N}^{12} \\
T_{0N}^{21} & T_{0N}^{22}
\end{pmatrix}
\begin{pmatrix}
E_F(x_{N-1}^{+}) \\
0
\end{pmatrix}.
\end{equation}
The overall transmission coefficient of the sample, denoted as \( t \), can be calculated as:

\begin{equation}
t =
\frac{E_F(x_{N-1}^{+})}{E_F(x_{0}^{-})} = \frac{1}{T_{0N}^{11}}.
\end{equation}
This transmission coefficient assumes we have the entire time domain signal. To disregard thick layers where the time delay associated with back-and-forth propagation of a wave excludes it from our measurement, we modify the transfer matrix to exclude the backward traveling wave from the thick layer to the previous layer is zero. While it is not possible to set \( E_B(x_{i}^+) \) to zero in equation \eqref{eq:Tij}, since this would invalidate the TMM approach, we can modify the individual transfer matrix elements in a way that is mathematically equivalent. To do this, we replace the second column elements in the transfer matrix \( T_{ij} \) with zero to form a modified one, \( M_{ij} \), as follows:
\begin{equation}
M_{ij} = 
\begin{pmatrix}
\frac{n_i + n_j}{2 n_i} & 0 \\
\frac{n_i - n_j}{2 n_i} & 0
\end{pmatrix}.
\end{equation}
This is valid, and has the effect of setting \( E_B(x_{i}^+) = 0 \) without interrupting subsequent calculations of wave propagation. The overall transfer function is then calculated in the usual way with the modified elements as appropriate. For example, in Fig.~\ref{fgr:fig1}, since layers 2 and 4 are thick, we will replace them with \( M_{12} \) and \( M_{34} \) as follows:

\begin{equation}
TF_{mod} = T_{0N} = T_{01} \cdot T_1 \cdot M_{12} \cdot T_2 \cdot T_{23} \cdot T_3 \cdot M_{34} \cdots T_{(N-1)} \cdot T_{(N-1)N}.
\end{equation}

\subsection{Fitting Techniques and Determining the Correct Dispersion Curve}

We employ a genetic algorithm (GA) to fit the modeled transfer function \( TF_{\text{mod}}(\omega, n_s) \) to the experimentally measured transfer function \( TF_{\text{exp}}(\omega) \). Through this optimization process, we determine the complex refractive index \( n_s \) by minimizing the cost function defined as:
\begin{equation}
CF = \left| TF_{\text{exp}}(\omega) - TF_{\text{mod}}(\omega, n_s) \right|.
\end{equation}
Before initiating the GA, we use an analytical approximation to generate initial estimates for the complex refractive index and set the lower and upper bounds of the optimization space. The former facilitates faster convergence toward the global minimum, while the latter ensures that only physically meaningful refractive indices are extracted. Since the GA is a metaheuristic optimization method that employs a population-based rather than single-point search, it is capable of exploring a broad solution space that does not depend on the initial population. Furthermore, since it avoids gradient information and the use of stochastic operations such as mutation and crossover, the algorithm can effectively escape local minima during the optimization process.
We initialize our optimization code using an analytic method that ignores multiple reflections and assumes that \( n \gg k \). In this limit, all terms in the transfer function cancel except those related to the layer that differs between the sample and reference.  Furthermore, it allows separation of the real and imaginary parts, enabling the analytical determination of both \( n \) and \( k \).  
Within these assumptions, the initial estimates for \( n(\omega) \) and \( k(\omega) \) are given by the following equations \cite{withayachumnankul2014fundamentals}:

\begin{equation}
n(\omega) = n_{r} + \frac{c\Delta \varphi(\omega)}{\omega d},
\label{eq:n_omega}
\end{equation}

\begin{equation}
k(\omega) = \frac{c}{\omega d} \cdot \ln \left( \frac{4n(\omega)n_{r}}{|TF_{\text{exp}}(\omega)| \cdot (n(\omega) + n_{r})^2} \right),
\label{eq:k_omega}
\end{equation}
where \( |TF_{\text{exp}}(\omega)| \) and \( \varphi(\omega) \) represent the amplitude and phase of the experimental transfer function, respectively. Since the phase is computed as the arctangent of the ratio of the imaginary to the real part of the transfer function, \( \varphi(\omega) \) is confined to the range \( -\pi/2 \) to \( \pi/2 \). Consequently, proper phase unwrapping is necessary to obtain a continuous phase response\cite{jepsen2019phase}. The quantity $n_{r}$ is the refractive index of the relevant layer in the reference measurement. For a film on a substrate, the space occupied by the material of interest has $n_{r} = 1$ (air or vacuum). However, this value can change in liquid phase experiments, where $n_{r} = n_{solvent}$ or in pump-probe experiments, where $n_{r}$ corresponds to the material in its ground state. 

The algorithm then begins with a set of candidate dispersion curves—one of which corresponds to the analytical solution—and, over successive iterations, minimizes the CF, progressively refining the complex refractive index to achieve a better fit with the experimental transfer function. However, after optimization, uncertainty in the correct phase offset may lead to an artificial shift in the calculated refractive index \( n' \), expressed as:
\begin{equation}
n' = n_{r} + \frac{c \Delta \varphi}{\omega d} \pm m \cdot \frac{2\pi c}{\omega d},
\label{eq:n_omega_general}
\end{equation}
where the integer \( m \) accounts for the \( 2\pi \) phase ambiguity. To constrain the optimization to physically meaningful values, we set symmetric boundaries around the analytical solution:

\begin{equation}
lb = n_{r} + \frac{c \Delta \varphi}{\omega d} - \frac{\pi c}{\omega d}, \quad
ub = n_{r} + \frac{c \Delta \varphi}{\omega d} + \frac{\pi c}{\omega d}.
\label{eq:n_bounds}
\end{equation}
Since phase retrieval is inherently modulo-\(2\pi\), limiting the search space (i.e., \(\text{ub} - \text{lb} = \frac{2\pi c}{\omega d}\)) ensures the inclusion of at least one solution.  The optimization process terminates when the stopping criterion is met—either the maximum number of iterations is reached, or the average relative change in the CF over several consecutive iterations falls below a specified function tolerance. 

\section{Results and discussion}
\textcolor{black}{To assess the performance and validate the accuracy of the MTMM method, we examine several sample structures with varying absorption properties, arbitrary thicknesses, and distinct configurations. We present results on several common sample types using previously published or newly measured data. These include a slab of a PA6 polymer material~\cite{d2014ultra}, conducting films deposited on a thick substrate~\cite{liu2023ultrasensitive, tayvah2021nelly}, and a cuvette filled with a water ~\cite{tayvah2021nelly}. The thicknesses of the layers in these samples can vary significantly; for instance, deposited layers span tens of nanometers to micrometers, slabs and liquid layers (within a cuvette) span tens to hundreds of micrometers, and substrates or cuvette walls are typically in the millimeter range. This variation indicates the practicality of this method in determining which layers fall within the temporal observation period and which do not. In some samples, additional complexity can be introduced using photoexcitation, which modifies the effective refractive index of a material over a characteristic length \cite{tayvah2021nelly}. These chosen examples represent a good set of test cases to evaluate our methods.} Detailed structural parameters for each sample are provided in Table~\ref{tab:parameters}.

By analyzing the time traces \textcolor{black}{from the experimental data, we determine the time window} ($\Delta \tau$) and the time difference between the signals passing through the reference and the sample ($\Delta t$). Utilizing Eq.~\ref{eq:neff}, we then determine the effective refractive index (\(n_{\text{eff}}\)) based on the observed delay between the sample and the reference. \textcolor{black}{Following this, we compute $d_{\text{th}}$ using Eq.~\ref{eq:dT} to determine which layers require a modified transfer matrix to exclude multiple reflections within them.} Table \ref{tab:parameters} presents the analytically calculated parameters for the three sample structures, with layers that exceed the threshold highlighted in orange.
\begin{table}[H]
    \centering
    \caption{Details of sample structure and extracted analytical values from time trace measurements.  In the sample Structure row, modified matrix elements are denoted using orange text, with unmodified elements in green.} 

    \begin{tabular}{|c|c|c|c|c|}
        \hline
        \diagbox[dir=NW]{\textbf{Parameters}}{\textbf{Sample}} 
            & \textbf{Cuvette} 
            & \textbf{Drop-Cast Film} 
            & \textbf{Slab} 
            & \textbf{PE Film} \\ \hline

        Sample Structure 
            & \textcolor{orange}{Glass} 
            & \textcolor{green2}{CP} 
            & \textcolor{green2}{PA6} 
            & \textcolor{green2}{SnO$_2^*$} \\

            & \textcolor{green2}{Water} 
            & \textcolor{orange}{Substrate} 
            & 
            & \textcolor{green2}{SnO$_2$} \\

            & \textcolor{orange}{Glass} 
            &  
            & 
            & \textcolor{orange}{Substrate} \\ \hline

        $\Delta t$ (fs) 
            & 367 
            & 35 
            & 730 
            & 273 \\ \hline

        $n_{\text{eff}}$ 
            & 2.1 
            & 2.5 
            & 1.47 
            & 82 \\ \hline

        $\Delta \tau$ (ps) 
            & 11 
            & 6 
            & 7.5 
            & 6.49 \\ \hline
    \end{tabular}
    \label{tab:parameters}
\end{table}

\begin{figure}[H]
    \centering
\includegraphics[width=1\linewidth]{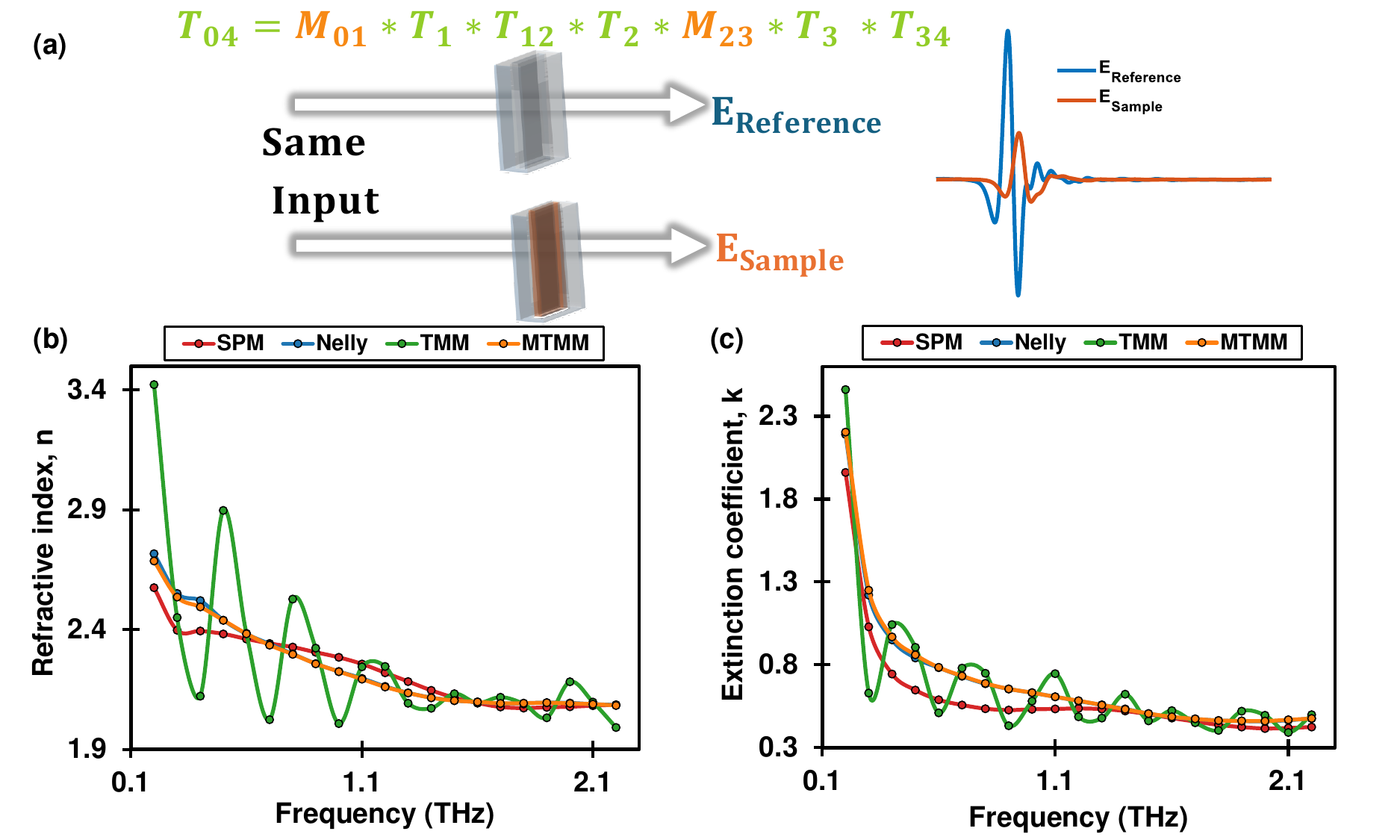}    
\caption{(a) Illustration of THz wave propagation through an empty cuvette (reference) and through a cuvette containing water (sample), with the signal observed within the first 11 ps time window. At the top of the schematic, the modified transfer function for the MTMM method is presented, where all transfer functions remain unchanged except for the ones highlighted in orange. (b) Refractive indices and (c) extinction coefficients extracted using SPM, MTMM, TMM, and the Nelly method. 
}
  \label{fgr:cuvette}
\end{figure}

We first apply our method to the example of a cuvette filled with water (Table~\ref{tab:parameters}) that is illustrated in Fig.~\ref{fgr:cuvette} (adapted from~\cite{tayvah2021nelly}). Signals from the both empty and filled cuvette are measured in the time domain, with the signal from the water-filled sample attenuated and delayed compared to the reference signal. Our analysis indicates that the cuvette’s outer glass layer (1.25 mm) exceeds the 832~$\mu$m threshold, whereas the internal water layer (100~$\mu$m) falls below the 793~$\mu$m threshold determined by Eq.~\ref{eq:dT}. Consequently, \textcolor{black}{as shown in the top panel of the Fig.~\ref{fgr:cuvette}a}, only \( M_{01} \) and \( M_{23} \) are modified, corresponding to the interface between media 0 and 1 and between 2 and 3. For other terms, we use the standard transfer matrices. We compare the calculated complex refractive index usng our MTMM method to other common approaches, including the full TMM, the approximate SPM, and Nelly tree method (Fig.~\ref{fgr:cuvette}b and c, respectively). The Nelly method is used as a benchmark due to its prior validation against finite-element simulations \cite{tayvah2021nelly}. While all methods show a general decreasing trend in complex refractive index with frequency, the TMM displays significant fluctuations and abrupt jumps in both the real and imaginary parts. This occurs because the TMM considers the entire transmitted signal, not just the portion within the time window, preventing the cost function from converging. The SPM provides a good approximation at higher frequencies; however, at lower frequencies, it deviates from the actual refractive index because it does not account for multiple reflections within the water sample. In contrast, our MTMM method yields highly accurate results across the entire range by accounting for multiple reflections within the water sample while disregarding reflections inside the thick glass material. 

\begin{figure}[h!]
    \centering
\includegraphics[width=1\linewidth]{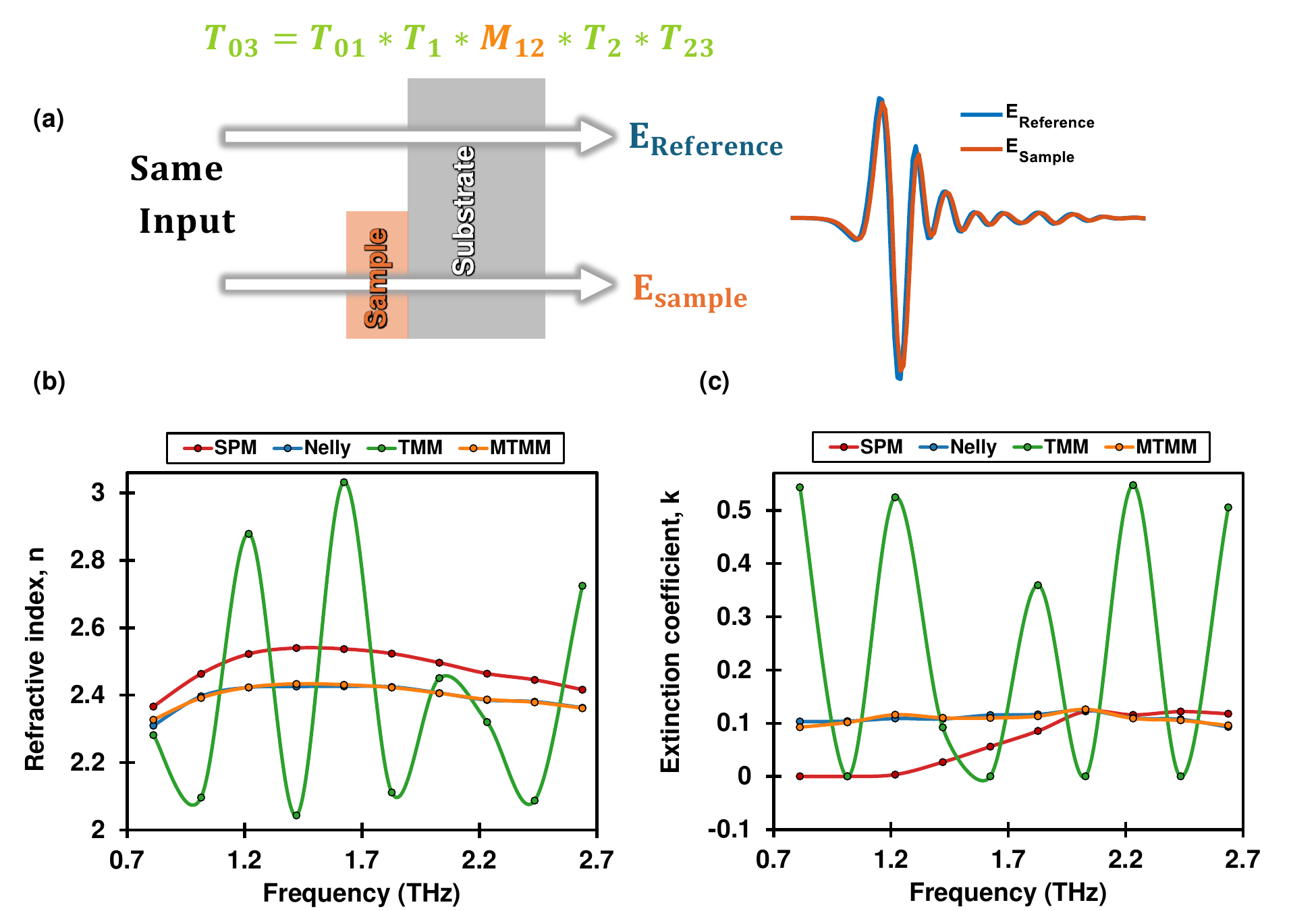}    
\caption{(a) Schematic representation of THz wave propagation through a sample consisting of a bare glass substrate on one half (reference) and a drop-cast layer of CF material on the other half. At the top of the schematic, the modified transfer function from the MTMM method is shown, with all other transfer matrices remaining unchanged except for $M_{12}$. (b) Refractive indices and (c) extinction coefficients obtained using the MTMM, SPM, and TMM methods, with cost functions of \(2.66 \times 10^{-7}\), \(0.013\), and \(0.2976\), respectively.}
  \label{fgr:thinfilm}
\end{figure}

We extend our investigation of the method by analyzing the time data obtained from a THz-TDS measurement of a thin film of a conducting polymer (CP) \cite{liu2023ultrasensitive} drop-cast on a glass substrate (Fig.~\ref{fgr:thinfilm}). The 7 micrometer sample thickness results in a time delay of only approximately 35 fs, significantly less than the 367~fs delay observed in the 100 micrometer cuvette sample. The drop-cast film’s CF dye (7 $\mu$m) is smaller than the calculated 332 $\mu$m threshold, but the substrate glass (500 $\mu$m) exceeds the calcualted 429 $\mu$m threshold. Therefore, as illustrated in the top panel of Fig.~\ref{fgr:thinfilm}a, only \( M_{12} \) is modified (at the interface between media 1 and 2), allowing the standard transfer matrix to be applied to the other layers. The complex refractive index of the material are calculated using the SPM, TMM, and MTMM approximation methods, as shown in Fig.~\ref{fgr:cuvette}b and c, respectively. Here, use the Nelly method as a benchmark to evaluate how closely each method converges to the experimental transfer function. The MTMM, SPM, and TMM methods yield cost functions of \(2.66 \times 10^{-7}\), \(0.013\), and \(0.2976\), respectively. This demonstrates the effectiveness of the MTMM compared to the other two methods. Again, the TMM shows significant fluctuations and fails to converge, while the SPM offers a reasonable approximation at higher frequencies but does not converge accurately at lower frequencies. In contrast, the MTMM method, by accounting for multiple reflections within the CF sample while disregarding reflections from the substrate, yields highly accurate results across the entire range.

\begin{figure}[H]
    \centering
    \includegraphics[width=1\linewidth]{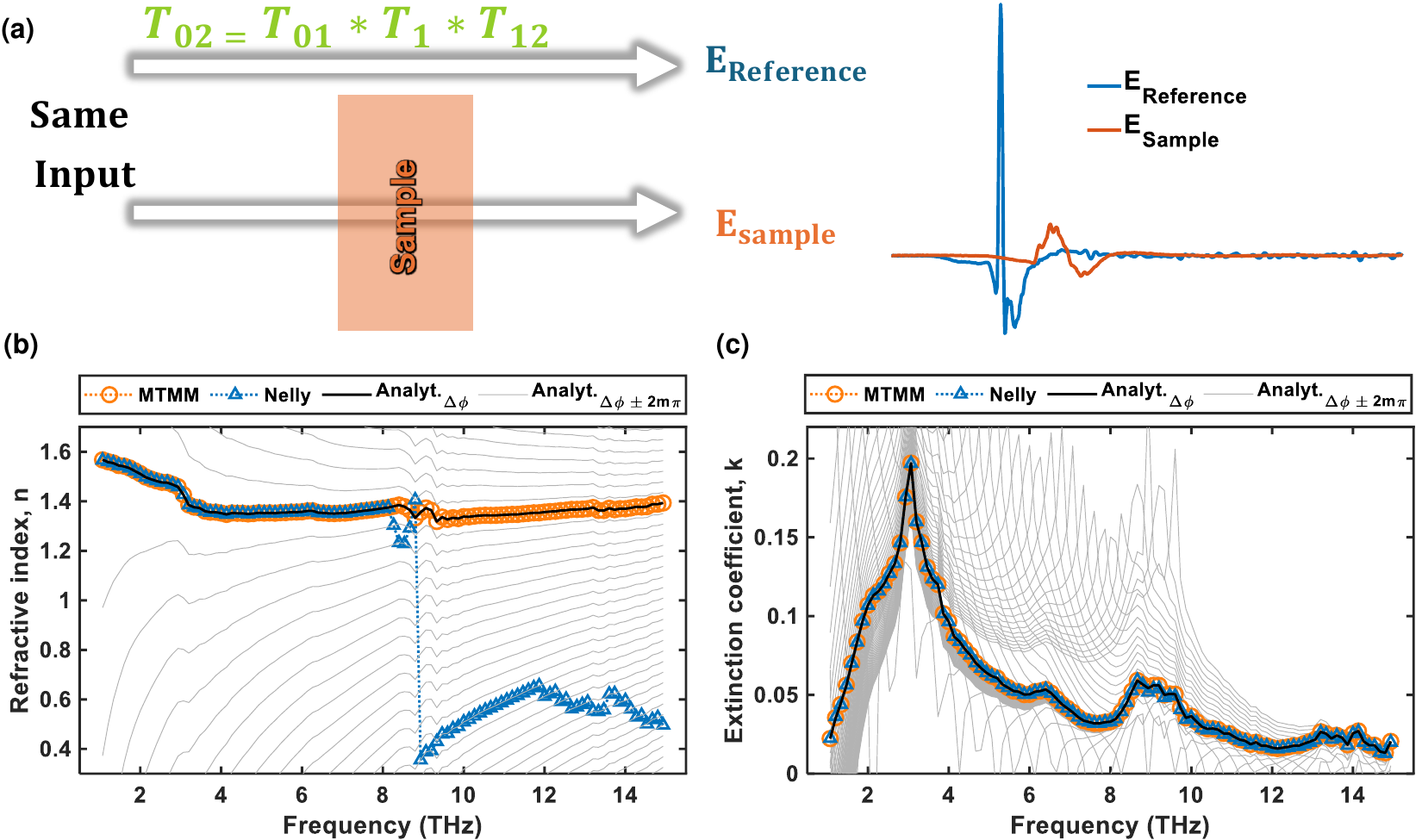}    
\caption{(a) Schematic of a broadband THz wave propagation through PA6 and air, with corresponding time-domain electric field intensities. (b) Refractive index and (c) extinction coefficient of PA6, extracted using the MTMM, Nelly, and SPM based analytical methods. Results are presented for both correctly unwrapped phase data and phase data offset by integer multiples of $\pm 2\pi$.}
    \label{fgr:PA6}
\end{figure}

We further investigate the time trace of a 470~$\mu$m thick polyamide-6 (PA6) slab, notable for its numerous absorption peaks across the 1–15~THz range (Fig.~\ref{fgr:PA6}), as reported in Francesco et al.~\cite{d2014ultra}. The PA6 slab (470 $\mu$m) is smaller than the 766 $\mu$m threshold. Therefore, we apply the full transfer matrix for this material, yielding identical results for both the MTMM and TMM methods. Fig.~\ref{fgr:PA6}b and c compare the calculated refractive index of PA6 using our MTMM model to other methods. Notably, the MTMM and single-pass analytical method (black curves in \ref{fgr:PA6}) closely agree with each other across the entire range. While the analytical method does not provide an accurate approximation for THz-TDS measurements in general, it does in this case since the sample consists of a single layer with a refractive index ($\sim$1.47) close to air ($\sim$1). As a result, the magnitude of its reflections, which is proportional to the difference in refractive indices, are low. Additionally, the low absorption of the material contributes to a reliable approximation. 

In contrast to previous examples, the Nelly tree-based method is not used as a benchmark, since it has not been validated over this geometry and frequency range.  In fact, we find that our population-optimization approach gives distinct results compared to the simpler optimization scheme implemented in Nelly at frequencies above 8 THz, after which large discontinuous jumps in the real part of the refractive index are observed. This discontinuities can be readily explained if we run our MTMM calculations without constraints, such that no unique solution exists due to uncertainty ($\Delta\phi \pm 2\pi m$) in the phase (gray curves in Fig. \ref{fgr:PA6}). This issue arises using Nelly because the optimization variables lack proper constraints and depend on the initial condition. As a result, it converges to the same local minimum each time the calculation is run and gives a physically meaningless solution.  In contrast, after several runs of optimization, the GA method explores multiple regions of the search space through population diversity and randomness to either (i) converge to the physically meaningful complex refractive index when proper constraints are applied or (ii) identify all possible solutions for this problem when no constraints are given (Fig. \ref{fgr:PA6_SI}). In the latter case, the correct dispersion curve can ultimately be extracted when suitable analytical methods can be used to confirm the correct solution.  It is worth noting that, although the real part exhibits discontinuities—since the optimization treats the real and imaginary components separately—the imaginary part remains relatively consistent. We note that these phase discontinuities are not observed in the solutions to thinner samples, such as 30~$\mu$m polytetrafluoroethylene (PTFE) (Fig. \ref{fgr:PTFE}). Here, the spacing between branches with a $2\pi$ phase shift is sufficiently large such that local minima are avoided in Nelly. However, since the number of internal reflections increases, the SPM method is no longer accurate.

\begin{figure}[H]
    \centering
\includegraphics[width=1\linewidth]{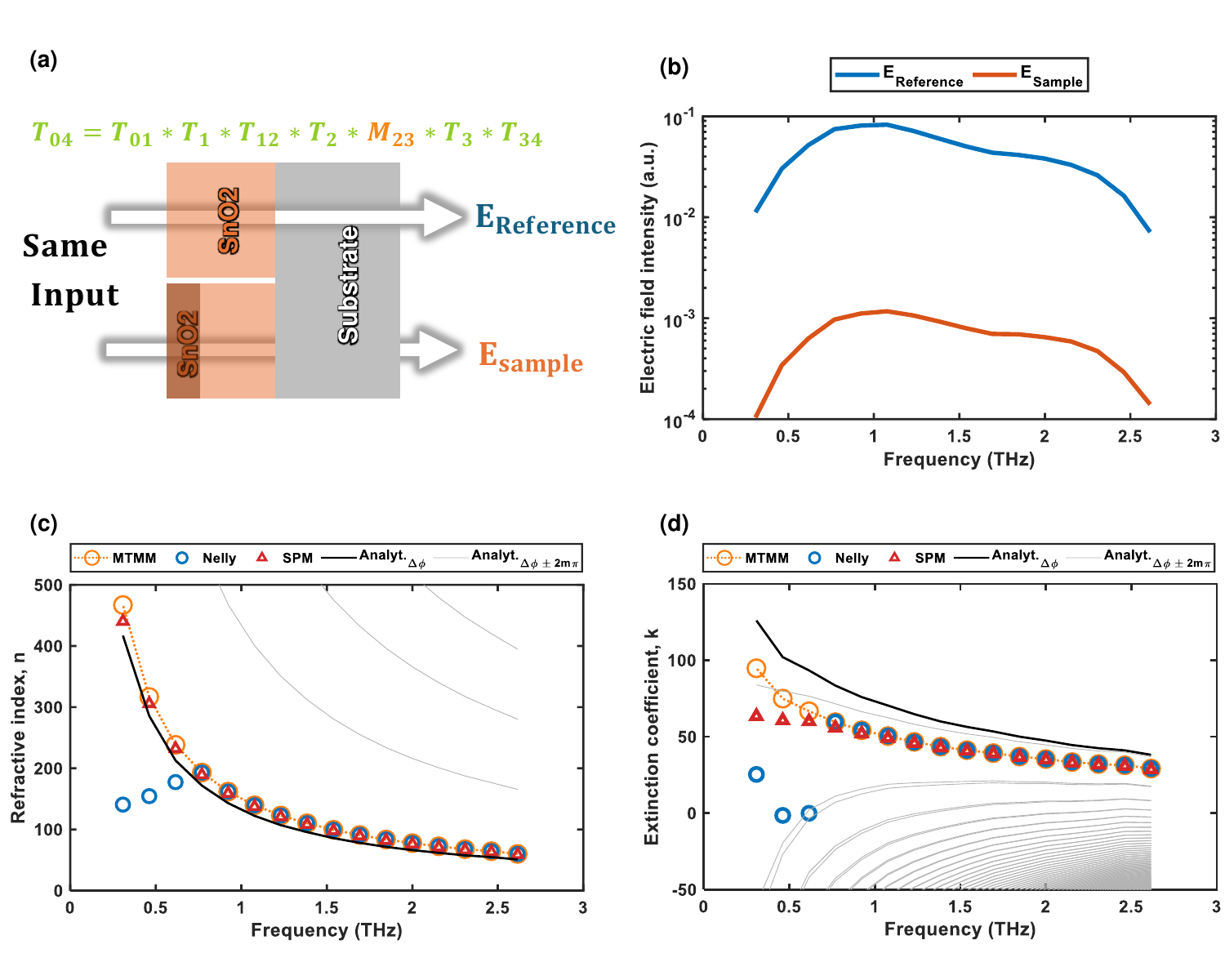}    
\caption{
(a) Illustration of the THz pulse propagating through an SnO\textsubscript{2} layer on a substrate, with and without photoexcitation. In this structure, only the glass substrate exceeds the thickness threshold. 
(b) Fourier-transformed spectra of the reference and sample signals, showing a significant reduction in intensity after photoexcitation. 
(c) Refractive indices and (d) extinction coefficients of the photoexcited SnO\textsubscript{2} layer, obtained using the TMM/MTMM, Nelly, and SPM methods, alongside analytical refractive indices for comparison.
}
  \label{fgr:SnO2}
\end{figure}

We note that our method can also be applied to model systems that contain photoexcited layers. Figure~\ref{fgr:SnO2}a illustrates a THz pulse propagating through an SnO\textsubscript{2} film with (sample) and without (reference) photoexcitation, based on experimental data from Tayvah et al.~\cite{tayvah2021nelly}. This structure consists of 1 $\mu$m of photoexcited SnO$_2$ and 7.22 $\mu$m of SnO$_2$, both well below estimated thickness thresholds, drop-cast onto glass layers (1 mm) that exceed the threshold. As such, a modified transfer matrix is only required for the substrate. After photoexcitation, the amplitude of the THz waveform is drastically attenuated compared to the reference (Fig.~\ref{fgr:SnO2}b), suggesting a high absorption coefficient in the excited state. We calculate the refractive index and extinction coefficient of the photoexcited SnO\textsubscript{2} layer using different methods (Fig.~\ref{fgr:SnO2}c,d). We observe good agreement between the SPM, Nelly, and MTMM methods above 0.7~THz. However, below 0.7~THz, the Nelly method fails and exhibits several discontinuities. There is no clear explanation for this behavior, as no corresponding jumps are observed in the Fourier spectrum of the data (Fig.~\ref{fgr:SnO2}b). This issue likely arises from limitations in the optimization algorithms, which may become trapped in local minima of the defined CF. As in other examples, the analytical SPM (black curves in Fig.~\ref{fgr:SnO2}c,d) also deviates from the MTMM method at lower frequencies due to the neglect of multiple reflections in the high-refractive-index material and the violation of the underlying assumption that the material exhibits very low absorption. However, a numerical SPM model employing GA optimization (with the same underlying transfer function as the analytical model) yields results that more closely match the actual values (red traingles in Fig.~\ref{fgr:SnO2}c,d), highlighting the advantage of numerical approaches for materials with large refractive indices.

\section{Conclusion}

In this work, we present a modified Transfer Matrix Method tailored for accurately extracting the refractive index in the THz regime. By carefully accounting for reflections and transmissions within a defined time window—and disregarding those beyond it—we achieved precise determination of material properties. The incorporation of a properly constrained genetic algorithm optimization enabled us to distinguish between physically meaningful solutions and mathematically valid but incorrect outcomes of the transfer function. We demonstrated the efficacy of this approach across various structures and material properties, such as a cuvette, a drop-cast film, a slab, and a photoexcited film, confirming its robustness in an ultrabroadband regime. These advancements pave the way for more reliable and efficient material characterization techniques in the THz domain, with potential applications in both fundamental research and advanced materials analysis. \textcolor{black}{Due to its effectiveness in calculating refraction and transmission, this method can also be extended to reflection measurements and other configurations.}

\section{Acknowledgments}
This work was supported by the Gordon and Betty Moore Foundation, grant 10.37807/gbmf12235. Aspects of this work were partially supported by the Designing Materials to Revolutionize and Engineer our Future (DMREF) program under grant agreement DMR-2323669. We gratefully acknowledge Mischa Bonn for providing the experimental PA6 and PTFE data used in this work. Additional data sets for the cuvette and photoexcited SnO\textsubscript{2} were adopted from \url{https://github.com/YaleTHz/nelly}.

\section{Disclosures}
The authors declare no conflicts of interest.

\section{Data availability}
Source code for the MTMM method and data files used in this manuscript are available at \url{https://github.com/SfeirLab/MTMM-THz-TDS}.

\bibliography{0_Reference}

\begin{thebibliography}{10}
\newcommand{\enquote}[1]{``#1''}

\bibitem{sun2017recent}
Q.~Sun, Y.~He, K.~Liu, \emph{et~al.}, \enquote{Recent advances in terahertz technology for biomedical applications,} {\protect\JournalTitle{Quantitative imaging in medicine and surgery}} \textbf{7}, 345 (2017).

\bibitem{stantchev2020real}
R.~I. Stantchev, X.~Yu, T.~Blu, and E.~Pickwell-MacPherson, \enquote{Real-time terahertz imaging with a single-pixel detector,} {\protect\JournalTitle{Nature communications}} \textbf{11}, 2535 (2020).

\bibitem{fischer2005chemical}
B.~Fischer, M.~Hoffmann, H.~Helm, \emph{et~al.}, \enquote{Chemical recognition in terahertz time-domain spectroscopy and imaging,} {\protect\JournalTitle{Semiconductor Science and Technology}} \textbf{20}, S246 (2005).

\bibitem{prokscha2025perspectives}
A.~Prokscha, F.~Sheikh, M.~Jalali, \emph{et~al.}, \enquote{Perspectives on terahertz honey bee sensing,} {\protect\JournalTitle{Scientific Reports}} \textbf{15}, 10638 (2025).

\bibitem{kleine2011review}
T.~Kleine-Ostmann and T.~Nagatsuma, \enquote{A review on terahertz communications research,} {\protect\JournalTitle{Journal of Infrared, Millimeter, and Terahertz Waves}} \textbf{32}, 143--171 (2011).

\bibitem{thorsmolle2004ultrafast}
V.~Thorsm{\o}lle, R.~Averitt, X.~Chi, \emph{et~al.}, \enquote{Ultrafast conductivity dynamics in pentacene probed using terahertz spectroscopy,} {\protect\JournalTitle{Applied physics letters}} \textbf{84}, 891--893 (2004).

\bibitem{yang2016biomedical}
X.~Yang, X.~Zhao, K.~Yang, \emph{et~al.}, \enquote{Biomedical applications of terahertz spectroscopy and imaging,} {\protect\JournalTitle{Trends in biotechnology}} \textbf{34}, 810--824 (2016).

\bibitem{seifert2016efficient}
T.~Seifert, S.~Jaiswal, U.~Martens, \emph{et~al.}, \enquote{Efficient metallic spintronic emitters of ultrabroadband terahertz radiation,} {\protect\JournalTitle{Nature photonics}} \textbf{10}, 483--488 (2016).

\bibitem{müller2020phase}
M.~Müller, N.~Mart{\'\i}n~Saban{\'e}s, T.~Kampfrath, and M.~Wolf, \enquote{Phase-resolved detection of ultrabroadband thz pulses inside a scanning tunneling microscope junction,} {\protect\JournalTitle{ACS photonics}} \textbf{7}, 2046--2055 (2020).

\bibitem{zhang2016broadband}
Y.~Zhang, X.~Zhang, S.~Li, \emph{et~al.}, \enquote{A broadband thz-tds system based on dstms emitter and ltg ingaas/inalas photoconductive antenna detector,} {\protect\JournalTitle{Scientific Reports}} \textbf{6}, 26949 (2016).

\bibitem{peretti2018thz}
R.~Peretti, S.~Mitryukovskiy, K.~Froberger, \emph{et~al.}, \enquote{Thz-tds time-trace analysis for the extraction of material and metamaterial parameters,} {\protect\JournalTitle{IEEE transactions on Terahertz Science and Technology}} \textbf{9}, 136--149 (2018).

\bibitem{ulatowski2020terahertz}
A.~M. Ulatowski, L.~M. Herz, and M.~B. Johnston, \enquote{Terahertz conductivity analysis for highly doped thin-film semiconductors,} {\protect\JournalTitle{Journal of Infrared, Millimeter, and Terahertz Waves}} \textbf{41}, 1431--1449 (2020).

\bibitem{Nuss1998}
M.~C. Nuss and J.~Orenstein, \emph{Terahertz time-domain spectroscopy} (Springer Berlin Heidelberg, Berlin, Heidelberg, 1998), pp. 7--50.

\bibitem{tayvah2021nelly}
U.~Tayvah, J.~A. Spies, J.~Neu, and C.~A. Schmuttenmaer, \enquote{Nelly: A user-friendly and open-source implementation of tree-based complex refractive index analysis for terahertz spectroscopy,} {\protect\JournalTitle{Analytical Chemistry}} \textbf{93}, 11243--11250 (2021).

\bibitem{jepsen2019phase}
P.~U. Jepsen, \enquote{Phase retrieval in terahertz time-domain measurements: a “how to” tutorial,} {\protect\JournalTitle{Journal of Infrared, Millimeter, and Terahertz Waves}} \textbf{40}, 395--411 (2019).

\bibitem{withayachumnankul2008uncertainty}
W.~Withayachumnankul, B.~M. Fischer, H.~Lin, and D.~Abbott, \enquote{Uncertainty in terahertz time-domain spectroscopy measurement,} {\protect\JournalTitle{Journal of the Optical Society of America B}} \textbf{25}, 1059--1072 (2008).

\bibitem{greenall2016multilayer}
N.~Greenall, L.~Li, E.~Linfield, \emph{et~al.}, \enquote{Multilayer extraction of complex refractive index in broadband transmission terahertz time-domain spectroscopy,} in \emph{2016 41st International Conference on Infrared, Millimeter, and Terahertz waves (IRMMW-THz),}  (IEEE, 2016), pp. 1--2.

\bibitem{cassar2019iterative}
Q.~Cassar, A.~Chopard, F.~Fauquet, \emph{et~al.}, \enquote{Iterative tree algorithm to evaluate terahertz signal contribution of specific optical paths within multilayered materials,} {\protect\JournalTitle{IEEE Transactions on Terahertz Science and Technology}} \textbf{9}, 684--694 (2019).

\bibitem{whelan2020reference}
P.~R. Whelan, Q.~Shen, D.~Luo, \emph{et~al.}, \enquote{Reference-free thz-tds conductivity analysis of thin conducting films,} {\protect\JournalTitle{Optics Express}} \textbf{28}, 28819--28830 (2020).

\bibitem{neu2018applicability}
J.~Neu, K.~P. Regan, J.~R. Swierk, and C.~A. Schmuttenmaer, \enquote{Applicability of the thin-film approximation in terahertz photoconductivity measurements,} {\protect\JournalTitle{Applied Physics Letters}} \textbf{113} (2018).

\bibitem{troparevsky2010transfer}
M.~C. Troparevsky, A.~S. Sabau, A.~R. Lupini, and Z.~Zhang, \enquote{Transfer-matrix formalism for the calculation of optical response in multilayer systems: from coherent to incoherent interference,} {\protect\JournalTitle{Optics express}} \textbf{18}, 24715--24721 (2010).

\bibitem{waddie2020terahertz}
A.~J. Waddie, P.~J. Schemmel, C.~Chalk, \emph{et~al.}, \enquote{Terahertz optical thickness and birefringence measurement for thermal barrier coating defect location,} {\protect\JournalTitle{Optics Express}} \textbf{28}, 31535--31552 (2020).

\bibitem{rashidi2024efficient}
K.~Rashidi, E.~Michail, B.~Salcido-Santacruz, \emph{et~al.}, \enquote{Efficient and tunable photochemical charge transfer via long-lived bloch surface wave polaritons,} {\protect\JournalTitle{arXiv preprint arXiv:2409.02067}}  (2024).

\bibitem{michail2024addressing}
E.~Michail, K.~Rashidi, B.~Liu, \emph{et~al.}, \enquote{Addressing the dark state problem in strongly coupled organic exciton-polariton systems,} {\protect\JournalTitle{Nano Letters}} \textbf{24}, 557--565 (2024).

\bibitem{katsidis2002general}
C.~C. Katsidis and D.~I. Siapkas, \enquote{General transfer-matrix method for optical multilayer systems with coherent, partially coherent, and incoherent interference,} {\protect\JournalTitle{Applied optics}} \textbf{41}, 3978--3987 (2002).

\bibitem{withayachumnankul2014fundamentals}
W.~Withayachumnankul and M.~Naftaly, \enquote{Fundamentals of measurement in terahertz time-domain spectroscopy,} {\protect\JournalTitle{Journal of Infrared, Millimeter, and Terahertz Waves}} \textbf{35}, 610--637 (2014).

\bibitem{wilk2008highly}
R.~Wilk, I.~Pupeza, R.~Cernat, and M.~Koch, \enquote{Highly accurate thz time-domain spectroscopy of multilayer structures,} {\protect\JournalTitle{IEEE Journal of Selected Topics in Quantum Electronics}} \textbf{14}, 392--398 (2008).

\bibitem{duvillaret1999highly}
L.~Duvillaret, F.~Garet, and J.-L. Coutaz, \enquote{Highly precise determination of optical constants and sample thickness in terahertz time-domain spectroscopy,} {\protect\JournalTitle{Applied optics}} \textbf{38}, 409--415 (1999).

\bibitem{ma2020application}
W.~Ma, C.~Li, Z.~Wang, \emph{et~al.}, \enquote{Application of terahertz time-domain spectroscopy in characterizing thin metal film--substrate structures,} {\protect\JournalTitle{IEEE Transactions on Terahertz Science and Technology}} \textbf{10}, 593--598 (2020).

\bibitem{lagarias1998convergence}
J.~C. Lagarias, J.~A. Reeds, M.~H. Wright, and P.~E. Wright, \enquote{Convergence properties of the nelder--mead simplex method in low dimensions,} {\protect\JournalTitle{SIAM Journal on optimization}} \textbf{9}, 112--147 (1998).

\bibitem{pupeza2007highly}
I.~Pupeza, R.~Wilk, and M.~Koch, \enquote{Highly accurate optical material parameter determination with thz time-domain spectroscopy,} {\protect\JournalTitle{Optics express}} \textbf{15}, 4335--4350 (2007).

\bibitem{chehouri2015review}
A.~Chehouri, R.~Younes, A.~Ilinca, and J.~Perron, \enquote{Review of performance optimization techniques applied to wind turbines,} {\protect\JournalTitle{Applied Energy}} \textbf{142}, 361--388 (2015).

\bibitem{klokkou2022artificial}
N.~Klokkou, J.~Gorecki, J.~S. Wilkinson, and V.~Apostolopoulos, \enquote{Artificial neural networks for material parameter extraction in terahertz time-domain spectroscopy,} {\protect\JournalTitle{Optics Express}} \textbf{30}, 15583--15595 (2022).

\bibitem{mirjalili2014grey}
S.~Mirjalili, S.~M. Mirjalili, and A.~Lewis, \enquote{Grey wolf optimizer,} {\protect\JournalTitle{Advances in engineering software}} \textbf{69}, 46--61 (2014).

\bibitem{Beddoes25}
B.~Beddoes, N.~Klokkou, J.~Gorecki, \emph{et~al.}, \enquote{Thz-tds: extracting complex conductivity of two-dimensional materials via neural networks trained on synthetic and experimental data,} {\protect\JournalTitle{Opt. Express}} \textbf{33}, 14872--14884 (2025).

\bibitem{kennedy1995particle}
J.~Kennedy and R.~Eberhart, \enquote{Particle swarm optimization,} in \emph{Proceedings of ICNN'95-international conference on neural networks,}  vol.~4 (ieee, 1995), pp. 1942--1948.

\bibitem{rashidi2023design}
K.~Rashidi, D.~Fathi, J.~Maleki, \emph{et~al.}, \enquote{Design of single-mode single-polarization large-mode-area multicore fibers,} {\protect\JournalTitle{Micromachines}} \textbf{14}, 1901 (2023).

\bibitem{rashidi2018optimal}
K.~Rashidi, S.~M. Mirjalili, H.~Taleb, and D.~Fathi, \enquote{Optimal design of large mode area photonic crystal fibers using a multiobjective gray wolf optimization technique,} {\protect\JournalTitle{Journal of Lightwave Technology}} \textbf{36}, 5626--5632 (2018).

\bibitem{yeh1990optical}
P.~Yeh and M.~Hendry, \enquote{Optical waves in layered media,}  (1990).

\bibitem{d2014ultra}
F.~D’Angelo, Z.~Mics, M.~Bonn, and D.~Turchinovich, \enquote{Ultra-broadband thz time-domain spectroscopy of common polymers using thz air photonics,} {\protect\JournalTitle{Optics express}} \textbf{22}, 12475--12485 (2014).

\bibitem{liu2023ultrasensitive}
C.-T. Liu, J.~Vella, N.~Eedugurala, \emph{et~al.}, \enquote{Ultrasensitive room temperature infrared photodetection using a narrow bandgap conjugated polymer,} {\protect\JournalTitle{Advanced Science}} \textbf{10}, 2304077 (2023).

\end{thebibliography}






\end{document}


\maketitle

\section{Extraction of refractive index in broadband and thick samples}

\begin{figure}
    \centering
\includegraphics[width=1\linewidth]{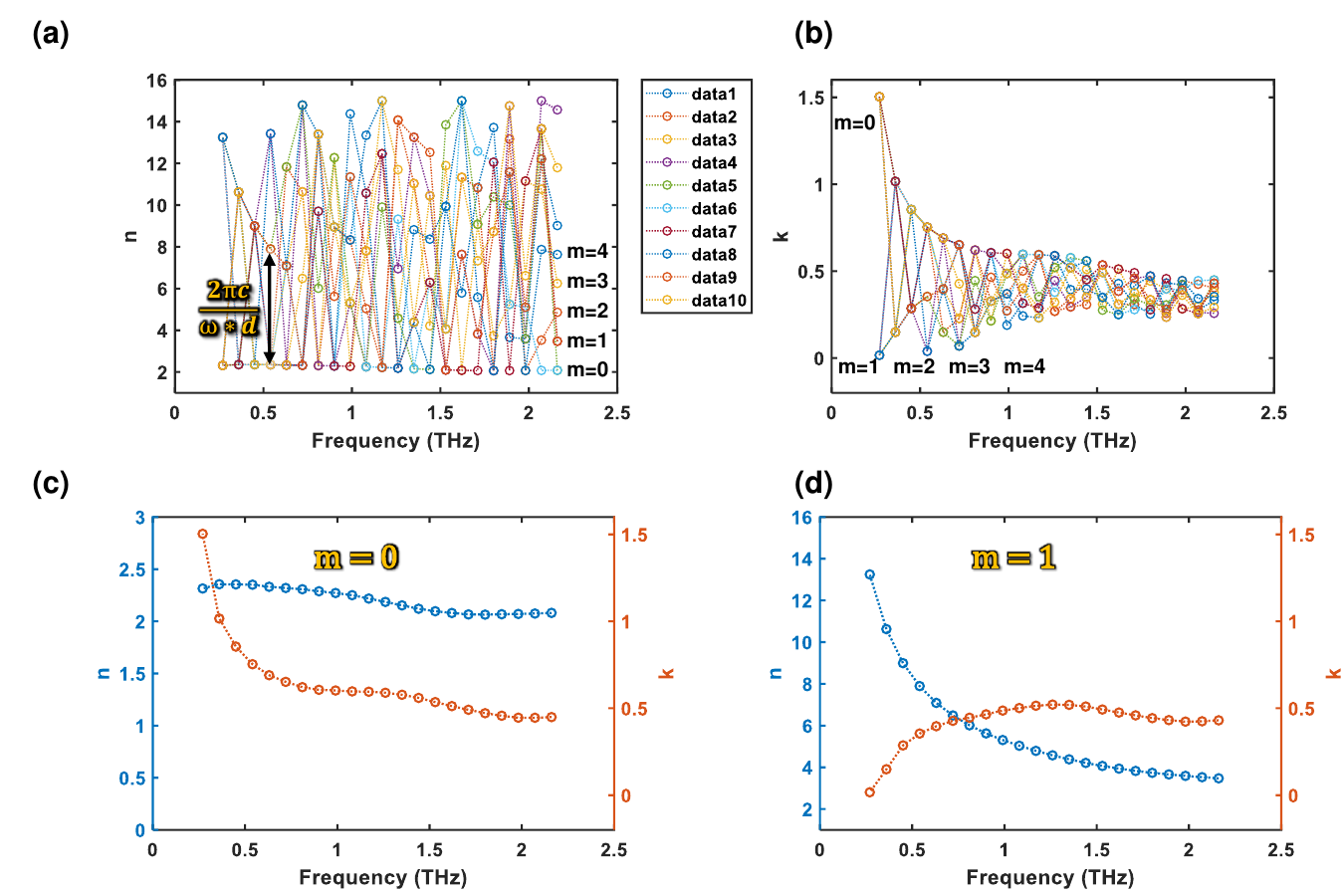} \caption{Results of 10-time optimization of the cuvette with water: (a) Refractive index and (b) extinction coefficient, all of which match the experimental transfer function. However, only one set corresponds to the true refractive index; the others differ by integer multiples of $\pm 2\pi$. (c) The first set represents the actual refractive index of the material, while (d) the second set corresponds to the first phase multiple but does not reflect the true refractive index.}
  \label{fgr:optimization}
\end{figure}

As illustrated in Eq.~14, the last term of the expression decreases with increasing $\omega$ or $d$, making it more difficult to distinguish between the physical and non-physical solutions for thicker samples or higher frequencies. Optimization algorithms must navigate these ambiguities, as shown in Fig.~\ref{fgr:optimization}, where repeated optimizations reveal results oscillating between the physical solution and various non-physical alternatives. 
While the goal is to cover a wider spectral range, higher frequencies reduce the distinction between physical and non-physical solutions. These issues are particularly pronounced in cuvettes with greater sample thicknesses (Fig.~2) and in broadband systems (Fig.~4). In contrast, drop-cast CP sample demonstrate fewer such ambiguities. Gradient-based optimization methods and the simplex method often converge to local minima, which may not represent the physical solution. In comparison, population-based optimization methods exhibit jumps between multiple integers of the phase in each optimization process, producing a broad set of both physical and non-physical results after several runs. By applying constraints through analytical methods, it becomes possible to isolate and recover the true refractive index.

\begin{figure}
    \centering    \includegraphics[width=0.8\linewidth]{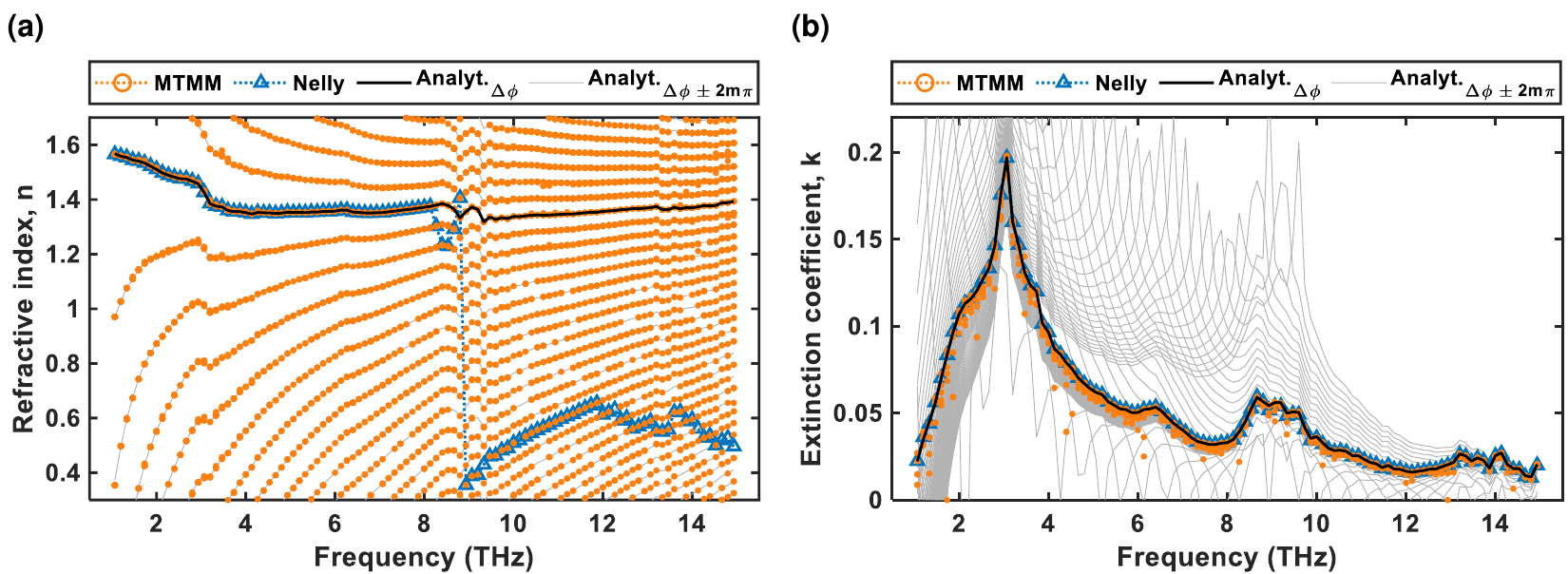}    \caption{(a) Refractive index and (b) extinction coefficient of PA6, extracted using the MTMM after multiple optimizations, the Nelly method, and SPM-based analytical approaches. Results are shown for both correctly unwrapped phase data and phase data offset by integer multiples of $\pm 2\pi$.}
    \label{fgr:PA6_SI}
\end{figure}

\begin{figure}
    \centering
\includegraphics[width=1\linewidth]{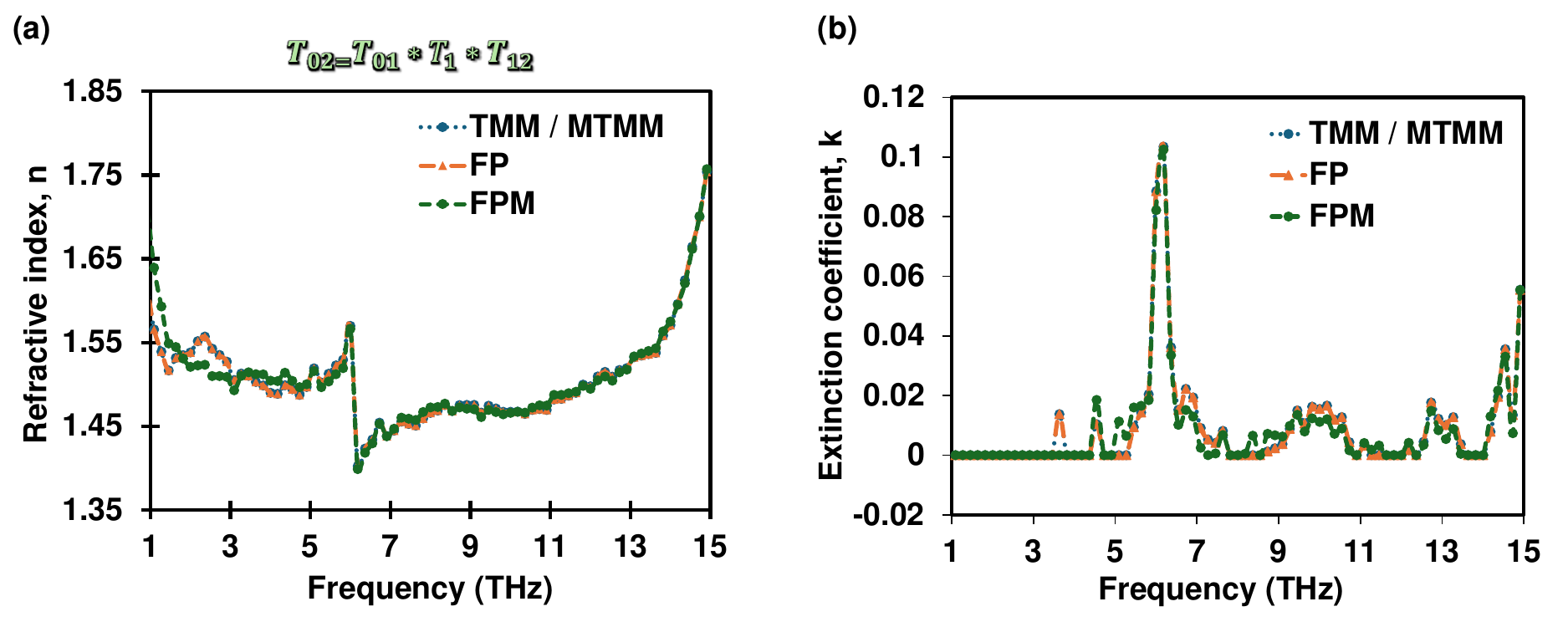}    
\caption{ (a) Refractive indices and (b) extinction coefficients of a thin (30 $\mu$m) PTFE sample obtained using the TMM/MTMM, Fabry-Perot (FP), and SPM methods. Since the sample thickness is low, the transfer function for TMM and MTMM are identical, as shown at the top of the figure, with all other transfer matrices remaining unchanged. The TMM/MTMM and FP methods yield identical results, while the SPM method differs from them.}
  \label{fgr:PTFE}
\end{figure}

\begin{figure}
    \centering
\includegraphics[width=1\linewidth]{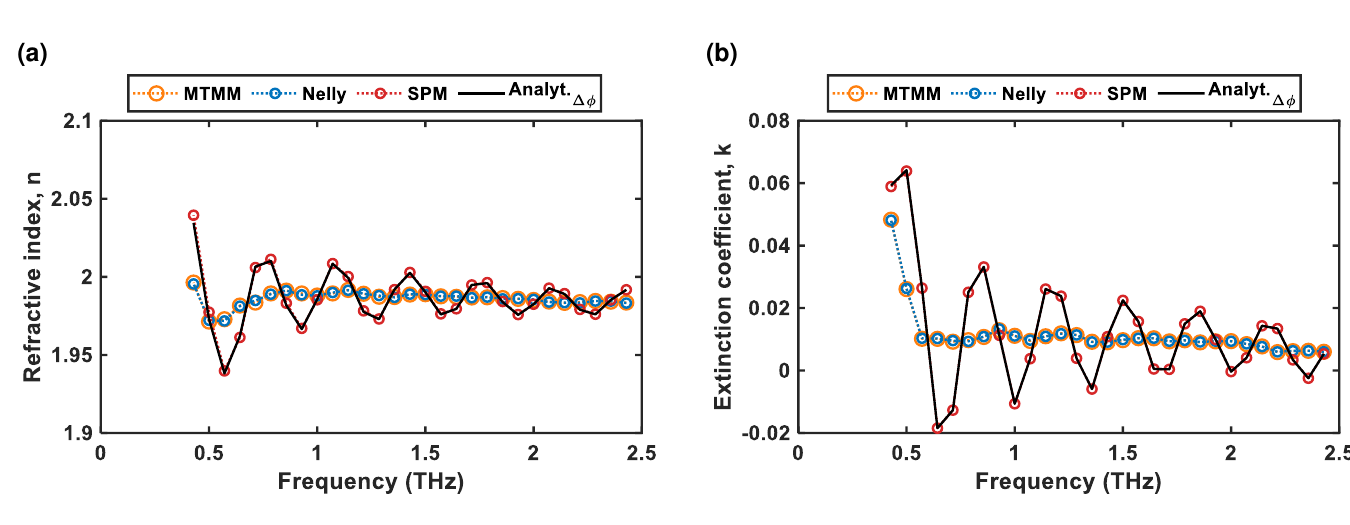}    
\caption{ (a) Refractive indices and (b) extinction coefficients of a thin (225 $\mu$m) quartz glass obtained using the TMM/MTMM, Nelly, SPM, and SPM based analytical methods. }
  \label{fgr:Quartz}
\end{figure}